\documentclass[12pt]{article}

\usepackage{amsmath,amssymb,amsfonts,amsthm,hyperref,bbm,latexsym,epsfig}
\usepackage{graphicx,multirow}
\usepackage{cite}

\newcommand{\bs}{\begin{subequations}}
\newcommand{\es}{\end{subequations}}
\newcommand{\be}{\begin{equation}}
\newcommand{\ee}{\end{equation}}
\newcommand{\ba}{\begin{eqnarray}}
\newcommand{\ea}{\end{eqnarray}}
\newcommand{\no}{\nonumber \\}

\newcommand{\zz}{\mathbbm{Z}}
\newcommand{\ie}{\textit{i.e.}}
\newcommand{\viz}{\textit{viz.}}
\newcommand{\cf}{\textit{cf.}}

\newcommand{\SG}{{\tt SmallGroups}}

\allowdisplaybreaks

\begin{document}

\title{
\normalsize \hfill CFTP/22-006
\\[6mm]
\LARGE The centers of discrete groups as stabilizers of Dark Matter}

\author{
\addtocounter{footnote}{2}
Darius Jur\v{c}iukonis,$^{(1)}$\thanks{E-mail:
  {\tt darius.jurciukonis@tfai.vu.lt}.}
  \
Lu\'\i s Lavoura$^{(2)}$\thanks{E-mail: {\tt balio@cftp.tecnico.ulisboa.pt}.}
\\*[3mm]
$^{(1)} \! $
\small University of Vilnius,
\small Institute of Theoretical Physics and Astronomy, \\
\small Saul\.{e}tekio av.~3, LT-10222 Vilnius, Lithuania 
\\[2mm]
$^{(2)} \! $
\small Universidade de Lisboa, Instituto Superior T\'ecnico, CFTP, \\
\small Av.~Rovisco Pais~1, 1049-001 Lisboa, Portugal
\\*[2mm]
}

\date{\today}

\maketitle

\begin{abstract}
  The most usual option to stabilize Dark Matter (DM) is a $\zz_2$ symmetry.
  In general,
  though,
  DM may be stabilized by any $\zz_N$ with $N \ge 2$.
  We consider the way $\zz_N$
  is a subgroup of the internal-symmetry group $G$ of a model;
  we entertain the possibility that $\zz_N$ is the center of $G$,
  yet $G$ is not of the form $\zz_N \times G^\prime$,
  where $G^\prime$ is a group smaller
  (\textit{i.e.}\ of lower order)
  than $G$.
  We examine all the discrete groups of order smaller than 2001
  and we find that many of them
  cannot be written as the direct product
  of a cyclic group and some other group,
  yet they have a non-trivial center
  that might be used in Model Building to stabilize DM.
\end{abstract}

\vspace*{4mm}

\section{Introduction}

The lightest Dark Matter (DM) particle ought to be stable
(\ie\ unable to decay),
or at least it should have a lifetime of order the age of the Universe.
If it is stable,
then there is an unbroken cyclic $\zz_N$ symmetry which is non-trivial
(\ie\ it has $N \ge 2$),
such that standard matter is invariant under $\zz_N$ while DM is not;
the $\zz_N$ charge different from $1$ of the lightest DM particle
prevents it from decaying to standard matter,
which has $\zz_N$ charge $1$.

The most usual option in Model Building is $N = 2$.
However,
some authors have considered possibilities $N > 2$.
For DM stabilized by a $\zz_3$ symmetry,
see Ref.~\cite{z3}.
Larger cyclic groups have been used to stabilize DM,
like $\zz_4$ and $\zz_6$~\cite{z4z6},
$\zz_5$~\cite{z5},
or a general $\zz_N$~\cite{zn}.

The $\zz_N$ that stabilizes DM may be the center of
a larger internal-symmetry group $G$.\footnote{The center of a group $G$
is its Abelian subgroup formed by the elements of $G$
that commute with all the elements of $G$.}
The simplest possibility consists in $G$ being a discrete group
of order\footnote{The order of a discrete group is the number
of its elements.}
$O$ that is isomorphic to the direct product $\zz_N \times G^\prime$,
where $G^\prime$ is a group of order $O / N$.
In that case,
all the irreducible representations (`irreps') of $G$
consist of the product of an irrep of $\zz_N$
(which is one-dimensional,
because $\zz_N$ is Abelian and Abelian groups have one-dimensional irreps)
and an irrep of $G^\prime$;
standard matter must be placed
in the trivial representation\footnote{The trivial representation of a group
is the one where all the group elements are mapped onto the unit matrix.}
of $\zz_N$ while DM is placed in non-trivial representations of $\zz_N$.

However,
also discrete groups $G$ that cannot be written
as the direct product of a cyclic group and a smaller group
may have a non-trivial $\zz_N$ center.
If that happens,
then once again an irrep of $G$ may represent $\zz_N$ either trivially
or in non-trivial fashion
(\viz\ when some elements of $\zz_N$ are represented
by a phase $f$ with $f \neq 1$ but $f^N = 1$).
If $\zz_N$ remains unbroken when $G$ is
(either softly or spontaneously)
broken,
and if there are particles with $\zz_N$ charge different from $1$,
then those particles play the role of DM,
while the particles with $\zz_N$ value $1$ are standard matter.

This possibility was recently called to our attention by Ref.~\cite{chulia},
where a group $G$ of order 81,
named $\Sigma(81)$,\footnote{The group $\Sigma(81)$ can \emph{not}
be written as a direct product $\zz_3 \times G^\prime$,
$G^\prime$ being a group of order 27.
Rather,
$\Sigma (81)$
(which has {\tt SmallGroups} identifier $\left[ 81,\, 7 \right]$)
is of the form $\left( \zz_3 \times \zz_3 \times \zz_3 \right) \rtimes \zz_3$,
\ie\ it is a semi-direct product.}\footnote{The group $\Sigma(81)$
was used in Model Building by E.~Ma\cite{ma}.
See also Ref.~\cite{ishimori}.}
was used as internal symmetry of a model.
The authors of Ref.~\cite{chulia} rightly pointed out that
``[some] irreducible representations [of $\Sigma(81)$]
form a closed set under tensor products,
implying that if every Standard Model field transforms as
[one of those representations],
then any field transforming as
[a representation that is not in that closed set]
will belong to the dark sector.
The lightest among them will then be a dark matter candidate.''

As a matter of fact,
this mechanism had already been suggested before,
\viz\ in Ref.~\cite{valle}.
There,
it was noted that some discrete subgroups of $SU(2)$
have subsets of irreps that are closed under tensor products,
and this fact might be used to stabilize DM.\footnote{In the course of
the present investigation we have found that this indeed happens
for \emph{all} the discrete subgroups of $SU(2)$, except the trivial subgroup.}

In this paper we make a survey of all the discrete groups
of order $O \le 2000$,
except groups with either $O = 512$,
$O = 1024$,
or $O = 1536$.
We select the groups that cannot be written as the direct product
of a non-trivial cyclic symmetry and a smaller group,
and that moreover have at least one \emph{faithful} irreducible representation
(`firrep').
We identify the center $\zz_N$ of each of those groups,
and also the dimensions $D$ of their firreps.
We construct various tables with the integers $O$,
$N$,
and $D$.
We find that very many discrete groups,
especially those that are not subgroups of any continuous group $SU(D)$,
have centers $\zz_N$ with $N \ge 2$,
and $N$ is sometimes quite large.

This paper is organized as follows.
In Sec.~2 we explain,
through the well-known cases of $SU(2)$ and $SU(3)$,
that some groups have a center $\zz_N$ with $N \ge 2$
and some irreps of those groups represent $\zz_N$ trivially
while other irreps do not.\footnote{Hurried readers may skip Section~2.}
In Sec.~3 we make a systematic survey of the centers
of all the discrete groups $G$ of order up to 2000
that cannot be written as the direct product
of a cyclic group and another group and that
have some faithful irreducible representation.\footnote{We do not survey
groups of order either~512,
1024,
or~1536,
because there are unpractically very many groups of those orders.}
In Sec.~4 we briefly state our conclusions.
As an Appendix to this paper,
comprehensive listings of the groups that we have studied
are available online at
\href{https://github.com/jurciukonis/GAP-group-search}{{\tt
    https://github.com/jurciukonis/GAP-group-search}}.

\section{$SU(2)$ and $SU(3)$}

\subsection{$SU(2)$}

The defining representation of $SU(2)$
consists of the $2 \times 2$ unitary matrices with determinant $1$.
One such matrix is
\be
\label{a2}
A_2 = \left( \begin{array}{cc} -1 & 0 \\ 0 & -1 \end{array} \right).
\ee
This is proportional to the unit matrix
and therefore it commutes with all the $2 \times 2$ matrices;
in particular,
it commutes with all the matrices in the defining representation of $SU(2)$.
Hence,
in any irrep of $SU(2)$,
$A_2$ must be represented by a matrix that commutes
with all the matrices in that irrep.
Schur's first lemma~\cite{elliott} states that
any matrix that commutes with all the matrices in an irrep of a group
must be proportional to the unit matrix.
Therefore,
in the $D$-dimensional irrep of $SU(2)$,\footnote{As is well known,
$SU(2)$ has one and only one $D$-dimensional irrep for each integer $D$.}
$A_2$ must be represented by a multiple of the $D \times D$ unit matrix
$\mathbbm{1}_D$.
But,
as is clear from Eq.~\eqref{a2},
$\left( A_2 \right)^2 = \mathbbm{1}_2$
is the unit element of $SU(2)$ in the defining representation.
This property must be reproduced in the $D$-dimensional irrep of $SU(2)$.
One hence concludes that,
in that irrep,
\be
A_2 \mapsto \left( - 1 \right)^{q_D} \times \mathbbm{1}_D,
\ee
where $q_D$ is an integer that is either 0 or 1 modulo 2.
The integer $q_D$ depends on the irrep.

The irreps of $SU(2)$ wherein $A_2$ is mapped onto the unit matrix,
\ie\ the ones for which $q_D$ is $0$ modulo 2,
are unfaithful.\footnote{An unfaithful representation of a group
represents two or more distinct elements by the same matrix.}
Those are the integer-spin irreps.
They have odd $D$ and are faithful irreps of the quotient group
\be
\label{so3}
SU(2) \left/ \ \zz_2 \right. \cong SO(3).
\ee
In Eq.~\eqref{so3},
\be
\label{z2}
\mathbbm{Z}_2 = \left\{ \mathbbm{1}_2,\ A_2 \right\}
\ee
is the center of $SU(2)$,
\textit{i.e.}\ it is the subset of $SU(2)$ elements
(in the defining representation)
that commute with \emph{all} the elements of $SU(2)$;
it is a $\mathbbm{Z}_2$ subgroup of $SU(2)$.

The $D$-dimensional irreps of $SU(2)$ with even $D$
are the half-integer-spin representations and represent $SU(2)$ faithfully,
\viz\ they map $A_2 \mapsto - \mathbbm{1}_D$.

Let us consider the tensor product of the irreps of $SU(2)$
with dimensions $D_1$ and $D_2$.
Clearly,
$A_2 \mapsto \left( -1 \right)^{q_{D_1}} \times \mathbbm{1}_{D_1}$
in the irrep with dimension $D_1$
and $A_2 \mapsto \left( -1 \right)^{q_{D_2}} \times \mathbbm{1}_{D_2}$
in the irrep with dimension $D_2$.
In the product representation,
which is in general reducible,
\be
\label{su2product}
A_2 \mapsto \left( -1 \right)^{q_{D_1} + q_{D_2}} \times \mathbbm{1}_{D_1 + D_2}.
\ee
Therefore,
the subset of the irreps of $SU(2)$ that have $q_D = 0$ modulo 2
is closed under tensor products.
This property of the irreps of $SU(2)$
also holds for the irreps of discrete subgroups of $SU(2)$.
If one such subgroup contains $A_2$ in its defining representation,
then a $D$-dimensional irrep of that subgroup
must represent $A_2$ either by $\mathbbm{1}_D$ or by $- \mathbbm{1}_D$,
and the subset of irreps that represent $A_2$ by unit matrices
is closed under tensor products.
It was suggested in Ref.~\cite{valle} that
this property may be used to stabilize DM.
In that suggestion,
Nature possesses an internal symmetry under a discrete subgroup of $SU(2)$
that contains in its defining irrep the matrix $A_2$;
standard matter sits in a (in general, reducible) representation
of that internal symmetry where $A_2$ is mapped onto the unit matrix,
while DM is in a representation of the internal symmetry
in which $A_2$ is mapped onto minus the unit matrix.
Then,
any collection of standard-matter particles will be invariant
under the transformation represented in the defining representation by $A_2$,
which implies that the lightest DM particle,
which changes sign under that transformation,
is stable.
Dark matter is stabilized by the $\zz_2$ symmetry in Eq.~\eqref{z2};
that $\zz_2$ symmetry is the center of the internal-symmetry group of the model.

\paragraph{Example:} The quaternion group $Q_8$
is the order-eight subgroup of $SU(2)$ formed,
in its defining two-dimensional irrep,
by the matrices\footnote{In Eqs.~\eqref{quat} and below,
we separate the classes of each group through semicolons.}
\bs
\label{quat}
\ba
& &
A = \left( \begin{array}{cc} i & 0 \\ 0 & -i \end{array} \right),
\qquad
A^3 = \left( \begin{array}{cc} -i & 0 \\ 0 & i \end{array} \right);
\\
& &
B = \left( \begin{array}{cc} 0 & 1 \\ -1 & 0 \end{array} \right),
\qquad
B^3 = \left( \begin{array}{cc} 0 & -1 \\ 1 & 0 \end{array} \right);
\\
& &
A B = \left( \begin{array}{cc} 0 & i \\ i & 0 \end{array} \right),
\qquad
B A = \left( \begin{array}{cc} 0 & -i \\ -i & 0 \end{array} \right);
\\
& &
A_2 = A^2 = B^2 = \left( \begin{array}{cc} -1 & 0 \\ 0 & -1 \end{array} \right);
\qquad
\left( A_2 \right)^2 = A^4 = B^4 =
\left( \begin{array}{cc} 1 & 0 \\ 0 & 1 \end{array} \right).
\label{ud}
\ea
\es
The center of this group consists of the $\zz_2$ in Eq.~\eqref{z2},
\cf\ line~\eqref{ud};
yet,
$Q_8$ is not the direct product of that $\zz_2$
and any order-four group.
The quaternion group has five irreps:
the two-dimensional one in Eqs.~\eqref{quat}
and the four one-dimensional ones
\be
\mathbf{1}_{rs}: \quad A \mapsto r,\ B \mapsto s,
\ee
where both $r$ and $s$ may be either $1$ or $-1$.
Clearly,
in all the one-dimensional irreps $A_2 = A^2 = B^2$
is mapped onto $1$,
while in the two-dimensional irrep $A_2$ is mapped onto $- \mathbbm{1}_2$.
In Model Building,
standard matter might sit in singlet irreps of $Q_8$
while DM would be placed in doublets of $Q_8$.
We envisage,
for instance,
an extension of the SM with \emph{global} symmetry $Q_8$
and four Higgs doublets $H_{1,2,3,4}$ that are singlets of $Q_8$ as
\be
H_1:\ \mathbf{1}_{++},\quad
H_2:\ \mathbf{1}_{+-},\quad
H_3:\ \mathbf{1}_{-+},\quad
H_4:\ \mathbf{1}_{--}.
\ee
If there are in the scalar potential quadratic terms $H_1^\dagger H_2$,
$H_1^\dagger H_3$,
$H_1^\dagger H_4$,
$H_2^\dagger H_3$,
$H_2^\dagger H_4$,
$H_3^\dagger H_4$,
and their Hermitian conjugates,
then the symmetry $Q_8$ is \emph{softly} broken---but its center $\zz_2$ is
preserved,
because $H_{1,2,3,4}$ are all invariant under it.
If either $H_2$,
$H_3$,
or $H_4$ acquire a VEV,
then the symmetry $Q_8$ is \emph{spontaneously} broken---but its center is,
once again,
preserved.
If additionally there is in the model some matter
(either fermionic or bosonic)
placed in doublets of $Q_8$,
then the lightest particle arising from that matter would be a DM candidate.

\subsection{$SU(3)$}

The defining representation of $SU(3)$
consists of the $3 \times 3$ unitary matrices with determinant $1$
and includes the matrix
\be
\label{a3}
A_3 = \left( \begin{array}{ccc}
  \omega & 0 & 0 \\ 0 & \omega & 0 \\ 0 & 0 & \omega \end{array} \right)
= \omega \times \mathbbm{1}_3,
\ee
where $\omega = \exp{\left( 2 i \pi / 3 \right)}$.
The Abelian group
\be
\label{z3}
\mathbbm{Z}_3 = \left\{ \mathbbm{1}_3,\ A_3,\ \left( A_3 \right)^2 \right\}.
\ee
forms the center of $SU(3)$ in the defining representation.
The matrix $A_3$ commutes with all the matrices in the defining representation
of $SU(3)$ and satisfies $\left( A_3 \right)^3 = \mathbbm{1}_3$.
Therefore,
in a $D$-dimensional irrep of $SU(3)$
\be
\label{qvalue}
A_3 \mapsto \omega^{q_D} \times \mathbbm{1}_D,
\ee
where $q_D$ is an integer that depends on the irrep and may be either $0$,
$1$,
or $2$ modulo $3$.\footnote{The value of $q_D$
is the `triality' of the irrep~\cite{khanna}.}
Irreps with $q_D = 0$
(like the octet and the decaplet)
have $A_3$ represented by $\mathbbm{1}_D$ and are unfaithful
representations of $SU(3)$.
Irreps with either $q_D = 1$
(like the triplet)
or $q_D = 2$
(like the sextet and the anti-triplet)
are faithful.
Clearly,
if $A_3 \mapsto \omega^{q_{D_1}} \times \mathbbm{1}_{D_1}$
in an irrep with dimension $D_1$
and $A_3 \mapsto \omega^{q_{D_2}} \times \mathbbm{1}_{D_2}$
in an irrep with dimension $D_2$,
then in the product representation
\be
\label{su3product}
A_3 \mapsto \omega^{q_{D_1} + q_{D_2}} \times \mathbbm{1}_{D_1 + D_2}.
\ee
Therefore,
the irreps with $q_D = 0$ form a closed set under tensor products.
There is a selection rule in tensor products of irreps of $SU(3)$,
similar to the selection rule in tensor products of irreps of $SU(2)$,
but with the group $\zz_3$ of Eq.~\eqref{z3} in $SU(3)$
instead of the group $\zz_2$ of Eq.~\eqref{z2} in $SU(2)$.

This also holds for many---but not all---the discrete subgroups of $SU(3)$.
The three matrices in Eq.~\eqref{z3}
may all belong to the defining representation
of a discrete subgroup of $SU(3)$;
when that happens,
a $D$-dimensional irrep of that subgroup possesses a $q_D$-value,
defined by Eq.~\eqref{qvalue}.
The $q_D$-values help determine the tensor products of irreps of the subgroup.
This may be used to explain the stability of DM:
if Nature had an internal symmetry that was a discrete subgroup of $SU(3)$
that contained the matrix $A_3$ in its defining representation
and that stayed unbroken,
then standard matter would sit in irreps of that subgroup with $q_D = 0$
while DM would be in irreps with either $q_D = 1$ or $q_D = 2$;
the lightest DM particle would then automatically be stable.

\paragraph{Example:} The group $A_4$ is the order-12 subgroup of $SU(3)$
formed,
in its defining representation,
by the matrices
\bs
\ba
& &
B = \left( \begin{array}{ccc}
  1 & 0 & 0 \\ 0 & -1 & 0 \\ 0 & 0 & -1 \end{array} \right),
\quad
A^2 B A = \left( \begin{array}{ccc}
  -1 & 0 & 0 \\ 0 & 1 & 0 \\ 0 & 0 & -1 \end{array} \right),
\quad
A B A^2 = \left( \begin{array}{ccc}
  -1 & 0 & 0 \\ 0 & -1 & 0 \\ 0 & 0 & 1 \end{array} \right);
\hspace*{7mm} \\
& &
A = \left( \begin{array}{ccc}
  0 & 1 & 0 \\ 0 & 0 & 1 \\ 1 & 0 & 0 \end{array} \right),
\quad
A B = \left( \begin{array}{ccc}
  0 & -1 & 0 \\ 0 & 0 & -1 \\ 1 & 0 & 0 \end{array} \right),
\no
& &
B A = \left( \begin{array}{ccc}
  0 & 1 & 0 \\ 0 & 0 & -1 \\ -1 & 0 & 0 \end{array} \right),
\quad
B A B = \left( \begin{array}{ccc}
  0 & -1 & 0 \\ 0 & 0 & 1 \\ -1 & 0 & 0 \end{array} \right);
\\
& &
A^2 = \left( \begin{array}{ccc}
  0 & 0 & 1 \\ 1 & 0 & 0 \\ 0 & 1 & 0 \end{array} \right),
\quad
A^2 B = \left( \begin{array}{ccc}
  0 & 0 & -1 \\ 1 & 0 & 0 \\ 0 & -1 & 0 \end{array} \right),
\no
& &
B A^2 = \left( \begin{array}{ccc}
  0 & 0 & 1 \\ -1 & 0 & 0 \\ 0 & -1 & 0 \end{array} \right),
\quad
A B A = \left( \begin{array}{ccc}
  0 & 0 & -1 \\ -1 & 0 & 0 \\ 0 & 1 & 0 \end{array} \right);
\quad
A^3 = B^2 = \left( \begin{array}{ccc}
  1 & 0 & 0 \\ 0 & 1 & 0 \\ 0 & 0 & 1 \end{array} \right).
\ea
\es
Neither the matrix $A_3$ nor $\left( A_3 \right)^2$
belong to the defining representation of $A_4$;
the center of $A_4$ is trivial,
\ie\ it is formed just by the unit element.
The four irreps $\mathbf{3}$,
$\mathbf{1}$,
$\mathbf{1}^\prime$,
and $\mathbf{1}^{\prime \prime}$ of $A_4$ do not have any selection rule
in their tensor products.
Thus,
the group $A_4$ is of no use to explain the stability of DM.

\section{Group search}

\subsection{Motivation}

The defining representation of $SU(D)$ consists of
the $D \times D$ unitary matrices with determinant $1$.
It is obvious that,
in this representation,
the center of $SU(D)$ is formed by the $D$ diagonal matrices
\be
\label{4jvio}
\Delta \times \mathbbm{1}_D, \
\Delta^2 \times \mathbbm{1}_D, \
\Delta^3 \times \mathbbm{1}_D, \
\ldots, \ \Delta^D \times \mathbbm{1}_D = \mathbbm{1}_D,
\ee
where $\Delta = \exp{\left( 2 i \pi / D \right)}$.
Thus,
the center of $SU(D)$ is a $\zz_D$ group.
Any discrete group that has a firrep
formed by matrices that belong to $SU(D)$
may contain in that representation either
\begin{description}
\item all the matrices in Eq.~\eqref{4jvio},
\item only the last one of them,
\item or---if $D$ is not a prime number and may be divided
  by an integer $m$ different from both $1$ and $D$---the
  $m^\mathrm{th}, 2 m^\mathrm{th}, \ldots, D^\mathrm{th}$
  matrices in Eq.~\eqref{4jvio}.
\end{description}
In general,
if $m$ is an integer that divides $D$
and $\mu = \exp{\left( 2 i \pi / m \right)}$,
then there is a cyclic symmetry $\zz_m$ given,
in the defining representation of $SU(D)$,
by
\be
\zz_m = \left\{ \mu \times \mathbbm{1}_D,\
\mu^2 \times \mathbbm{1}_D,\
\mu^3 \times \mathbbm{1}_D,\ \ldots,\
\mu^m \times \mathbbm{1}_D = \mathbbm{1}_D \right\}.
\ee
Some discrete subgroups of $SU(D)$ may then have $\zz_m$ as their center.

Thus,
discrete subgroups of $SU(D)$
that are not subgroups of any $U(D^\prime)$ with $D^\prime < D$
may have very few centers.
For instance,
a discrete subgroup of $SU(10)$
that is not a subgroup of any $U(D^\prime)$ with $D^\prime < 10$
may only have center $\zz_2$,
$\zz_5$,
$\zz_{10}$,
or the trivial group;
and a discrete subgroup of $SU(11)$
that is not a subgroup of any $U(D^\prime)$ with $D^\prime < 11$
may only have center $\zz_{11}$ or the trivial group.

\paragraph{Example:}
Consider the discrete group generated by two transformations $b$ and $c$
that obey
\be
c^8 = e,\quad b^4 = c^4,\quad c^2 b c^2 = b,\quad c^3 b = b^3 c,
\label{jgf9440}
\ee
where $e$ is the identity transformation.
There is a four-dimensional irreducible representation
of Eqs.~\eqref{jgf9440} as
\be
b \mapsto \left( \begin{array}{cccc}
  0 & 0 & 1 & 0 \\
  0 & 0 & 0 & -1 \\
  0 & 1 & 0 & 0 \\
  1 & 0 & 0 & 0
\end{array} \right),
\qquad
c \mapsto \left( \begin{array}{cccc}
  0 & 0 & 1 & 0 \\
  0 & 0 & 0 & 1 \\
  0 & 1 & 0 & 0 \\
  -1 & 0 & 0 & 0
\end{array} \right).
\label{jifspre}
\ee
Both matrices in Eqs.~\eqref{jifspre} are orthogonal and have determinant $+1$,
therefore this group~\cite{group327}
is a subgroup of both $SO(4)$ and $SU(4)$.
One easily sees that in the representation~\eqref{jifspre}
\be
b^4 \mapsto \mathrm{diag} \left( -1,\ -1,\ -1,\ -1 \right),
\label{vvfigtot}
\ee
while $b^2$ is not mapped onto a diagonal matrix.
Hence,
this subgroup of $SU(4)$ has center $\zz_2$ generated by $b^4$.
One finds that the defining conditions~\eqref{jgf9440} allow
two inequivalent doublet representations:
\be
\mathbf{2}_1: \qquad
b \mapsto \left( \begin{array}{cc} 1 & 0 \\ 0 & -1 \end{array} \right),
\quad
c \mapsto \left( \begin{array}{cc} 0 & -1 \\ 1 & 0 \end{array} \right),
\label{doisum}
\ee
and
\be
\mathbf{2}_2: \qquad
b \mapsto \left( \begin{array}{cc} 0 & -1 \\ 1 & 0 \end{array} \right),
\quad
c \mapsto \left( \begin{array}{cc} 1 & 0 \\ 0 & -1 \end{array} \right).
\label{doisdois}
\ee
Additionally,
the conditions~\eqref{jgf9440} have eight inequivalent singlet representations:
\be
\mathbf{1}_{p}:\ b \mapsto i^{\, p},\ c \mapsto i^{\, p}
\quad \mbox{and} \quad
\mathbf{1}_{4+p}:\ b \mapsto i^{\, p},\ c \mapsto - i^{\, p},
\quad \mbox{where} \quad
p \in \left\{ 0, 1, 2, 3 \right).
\label{uns}
\ee
In irreps~\eqref{doisum}--\eqref{uns}
the transformation $b^4$ is represented by the unit matrix
instead of minus the unit matrix as in Eq.~\eqref{vvfigtot};
therefore,
those irreps are unfaithful.
This group thus has
only one firrep---the quadruplet~\eqref{jifspre}---and ten unfaithful
inequivalent irreps---the two doublets~\eqref{doisum}
and~\eqref{doisdois} and the eight singlets~\eqref{uns}.
If one had a theory with this group as internal symmetry
and in that theory the only scalars that acquired VEVs
were placed in unfaithful irreps of the group,
then the internal symmetry
would get spontaneously broken to the $\zz_2$ generated by $b^4$.
Alternatively,
the theory might have soft-breaking terms
of types either $\mathbf{1}_p^\dagger \mathbf{1}_q$
(with $p \neq q$)
or $\mathbf{2}_1^\dagger \mathbf{2}_2$,
and then the discrete group would be broken softly
but its $\zz_2$ subgroup would remain unbroken.
Any fields in such a theory placed in quadruplets of the symmetry group
might then take the role of DM.

\vspace*{5mm}

Discrete subgroups of $U(D)$ do not bear the constraint that
the determinants of the matrices in their defining representations
should be $1$.
As a consequence,
if
\be
\zz_t = \left\{ \theta \times \mathbbm{1}_D,\
\theta^2 \times \mathbbm{1}_D,\
\theta^3 \times \mathbbm{1}_D,\ \ldots,\
\theta^t \times \mathbbm{1}_D = \mathbbm{1}_D \right\},
\ee
where $\theta = \exp{\left( 2 i \pi / t \right)}$,
is the center of a discrete subgroup of $U(D)$,
then there appears to be \textit{a priori}\/ no restriction on $t$.

\paragraph{Example:} The discrete group $\zz_8 \rtimes \zz_2$ has order 16
and \SG\ identifier $\left[ 16,\, 6 \right]$.
In its defining representation it is formed by the matrices
\bs
\label{v8r90}
\ba
& &
\left( \begin{array}{cc} 1 & 0 \\ 0 & -1 \end{array} \right),
\quad
\left( \begin{array}{cc} -1 & 0 \\ 0 & 1 \end{array} \right);
\quad
\left( \begin{array}{cc} i & 0 \\ 0 & -i \end{array} \right),
\quad
\left( \begin{array}{cc} -i & 0 \\ 0 & i \end{array} \right);
\\
& &
\left( \begin{array}{cc} 0 & -1 \\ i & 0 \end{array} \right),
\quad
\left( \begin{array}{cc} 0 & 1 \\ -i & 0 \end{array} \right);
\quad
\left( \begin{array}{cc} 0 & i \\ 1 & 0 \end{array} \right),
\quad
\left( \begin{array}{cc} 0 & -i \\ -1 & 0 \end{array} \right);
\\
& &
\left( \begin{array}{cc} 0 & i \\ -1 & 0 \end{array} \right),
\quad
\left( \begin{array}{cc} 0 & -i \\ 1 & 0 \end{array} \right);
\quad
\left( \begin{array}{cc} 0 & -1 \\ -i & 0 \end{array} \right),
\quad
\left( \begin{array}{cc} 0 & 1 \\ i & 0 \end{array} \right);
\\
& &
\left( \begin{array}{cc} i & 0 \\ 0 & i \end{array} \right);
\quad
\left( \begin{array}{cc} -i & 0 \\ 0 & -i \end{array} \right);
\quad
\left( \begin{array}{cc} -1 & 0 \\ 0 & -1 \end{array} \right);
\quad
\left( \begin{array}{cc} 1 & 0 \\ 0 & 1 \end{array} \right).
\label{8tg9u}
\ea
\es
This is the firrep $\mathbf{2}$ of $\zz_8 \rtimes \zz_2$.
The other inequivalent irreps of that group are the $\mathbf{2}^\ast$
(wherein each matrix of the $\mathbf{2}$
is mapped onto its complex-conjugate matrix)
and eight inequivalent unfaithful singlet irreps.
Most of the $2 \times 2$ unitary matrices~\eqref{v8r90}
do not have determinant $1$;
therefore,
$\zz_8 \rtimes \zz_2$ is a subgroup of $U(2)$ but not of $SU(2)$.
One sees in line~\eqref{8tg9u} that the center of $\zz_8 \rtimes \zz_2$ is
\be
\zz_4 = \left\{
\left( \begin{array}{cc} i & 0 \\ 0 & i \end{array} \right),\
\left( \begin{array}{cc} -1 & 0 \\ 0 & -1 \end{array} \right),\
\left( \begin{array}{cc} -i & 0 \\ 0 & -i \end{array} \right),\
\left( \begin{array}{cc} 1 & 0 \\ 0 & 1 \end{array} \right) \right\}.
\ee
Thus,
while discrete subgroups of $SU(2)$ may have center either $\zz_1$ or $\zz_2$,
discrete subgroups of $U(2)$ enjoy further possibilities,
for instance $\zz_4$.

Motivated by this observation that discrete subgroups of $U(D)$
may in general have diverse centers,
in our work we have surveyed many discrete groups
in order to find out their centers
and also which groups $U(D)$ they are subgroups of.

\subsection{{\tt GAP} and {\tt SmallGroups}}

{\tt GAP}~\cite{gap} is a computer algebra
that provides a programming language and
includes many functions that implement various algebraic algorithms.
It is supplemented by libraries
containing a large amount of data on algebraic objects.
With {\tt GAP} it is possible to study groups and their representations,
to display the character tables,
to find the subgroups of larger groups,
to identify groups given through their generating matrices,
and so on.

{\tt GAP} allows access to the {\tt SmallGroups} library~\cite{sg}.
This library contains all the finite groups of order less than 2001,
but for order 1024---because there are many thousands of millions of groups
of order 1024.
{\tt SmallGroups} also contains some groups for some specific orders
larger than 2000.
In {\tt SmallGroups} the groups are ordered by their orders;
for each order,
the complete list of nonisomorphic groups is given.
Each discrete group of order smaller than $2001$
is labeled $\left[ O,\, n \right]$ by {\tt SmallGroups},
where $O \le 2000$ is the order of the group
and $n \in \mathbb{N}$ is an integer that distinguishes
among the non-isomorphic groups of the same order.

\subsection{Procedure}

We have surveyed all the discrete groups of order $O \le 2000$
in the {\tt SmallGroups} library,
except the groups of order either 512,
1024,
or 1536.\footnote{Rather exceptionally,
we have included in our search four groups
of order 1536 that are known to have three-dimensional firreps,
according to our previous paper~\cite{jurciukonis}.}
We have discarded all the groups that are isomorphic to the direct product
of a smaller
(\ie\ of lower order) group and a cyclic group.\footnote{{\tt SmallGroups}
itself informs us about the structure of each group,
\viz\ whether it is isomorphic to the direct product of smaller groups.
We have found that there are,
however,
at least two exceptions.
One of them is the group with {\tt SmallGroups} identifier
$\left[ 180,\, 19 \right]$;
{\tt SmallGroups} informs us that
this is the group $\mathrm{GL} \left( 2,\, 4 \right)$
but omits the well-known fact that $\mathrm{GL} \left( 2,\, 4 \right)$
is isomorphic to $\zz_3 \times A_5$,
where $A_5$ is the group of the even permutations of five objects.
(Thus,
$\left[ 180,\, 19 \right]$ is discarded in our search,
because it is the direct product of $A_5$
and the cyclic group $\zz_3$.)
The other exception are the dihedral groups $D_O$ of order $O = 12 + 8 p$,
where $p$ is an integer,
\viz\ the groups $D_{12}$,
$D_{20}$,
$D_{28}$,
and so on.
({\tt SmallGroups} instead uses the notation $D_{O/2}$ for these groups,
\viz\ it uses $D_6$,
$D_{10}$,
$D_{14}$,
and so on.)
It is easy to check analytically that these specific $D_O$ groups
are isomorphic to $\zz_2 \times D_{O/2}$,
but {\tt SmallGroups} omits this fact.
We have used a method,
suggested to us by G\'abor Horv\'ath,
to check whether \emph{any} group $\left[ O,\, n \right]$
is a direct product of smaller groups,
namely the succession of {\tt GAP} commands
{\tt G := SmallGroup(O, n)}
and {\tt ListX(DirectFactorsOfGroup(G), StructureDescription)}.}
We have used {\tt GAP} to find out all the irreps of each remaining group,
and then to ascertain whether those irreps are faithful or not.
We have discarded all the groups
that do not have any firrep.\footnote{There are groups
for which all the faithful representations are reducible.
An example is the group formed by the 32 matrices
\be
\left( \begin{array}{cccc}
  a & 0 & 0 & 0 \\ 0 & b & 0 & 0 \\ 0 & 0 & c & 0 \\ 0 & 0 & 0 & d
\end{array} \right)
\quad \mbox{and}\, \quad
\left( \begin{array}{cccc}
  0 & a & 0 & 0 \\ b & 0 & 0 & 0 \\ 0 & 0 & 0 & c \\ 0 & 0 & d & 0
\end{array} \right),
\label{uit3}
\ee
where $a$,
$b$,
$c$,
and $d$ may be either $1$ or $-1$.
This group---with \SG\ identifier $\left[ 32,\, 27 \right]$
and structure $\left(
\zz_2 \times \zz_2 \times \zz_2 \times \zz_2 \right) \rtimes \zz_2$---has
eight inequivalent singlet irreps and six inequivalent doublet irreps,
but all of them are unfaithful.
The defining representation in Eq.~\eqref{uit3} is,
by definition,
faithful,
but it is reducible.}
We have thus obtained 87,349 non-isomorphic groups,
that are all listed in our tables available at the site 
\href{https://github.com/jurciukonis/GAP-group-search}{{\tt
    https://github.com/jurciukonis/GAP-group-search}}.
We have looked only at the firreps of each group;
non-faithful irreps,
and all reducible representations,
were neglected.
We have computed the determinants of the matrices of each firrep
in order to find out whether all those determinants are $1$ or not.
We have also looked for all the matrices in the firreps
that are proportional to the unit matrix
and we have checked that those matrices form a group $\zz_N$
for some integer $N$
(which for some groups is just $1$).

We have tried to answer the questions
on how the following three integers are related:
\begin{enumerate}
\item The integer $O$ that is the order of the discrete group $G$.
\item The integer $N$ corresponding to the group $\zz_N$
  that is the center of $G$.
\item The integer $D$ such that $G$ has one or more firreps of dimension $D$.
\end{enumerate}
We have also examined the question whether each $D$-dimensional firrep
is equivalent to a representation through matrices
of $SU(D)$.\footnote{All the irreps of discrete groups are equivalent
to representations through unitary matrices,
and therefore we know that the generators that {\tt GAP} provides to us
are equivalent to unitary generators,
even though {\tt GAP} often gives them in non-unitary version.
In order to know whether the generators belong to $SU(D)$
we just compute their determinants.}

There are relatively few groups that have firreps
with different dimensions.
(For instance:
$A_5$ has firreps of dimensions three,
four,
and five.
On the other hand,
the group $\Sigma \left( 36 \times 3 \right)$,
that has \SG\ identifier $\left[108,\, 15 \right]$,
has irreps of dimensions $1$,
$3$,
and $4$,
but the $1$- and $4$-dimensional irreps are unfaithful---all the firreps
have dimension $3$.)
We have found just 2787 such discrete groups,
out of the total 87,349 groups that we have surveyed;
they are collected in table Intersections at
\href{https://github.com/jurciukonis/GAP-group-search}{{\tt
    https://github.com/jurciukonis/GAP-group-search}}.

\paragraph{Computing time:}
The scan over the \SG\ library
to find the firreps of all possible dimensions
constituted a computationally very expensive task.
Our computations with {\tt GAP} took about three months.
It is difficult to estimate the total number of CPU hours (CPUH) spent
in the computations,
because various computers with different CPUs were used.
Most of the time was consumed in the computation of the irreps of the groups.
For example,
the computation for group $\left[ 1320,\, 15 \right]$,
\viz\ $\mathrm{SL} \left( 2, 11 \right)$,
took about 320~CPUH running on Intel Xeon~CPU~@~1.60GHz
or about 46~CPUH in the newer Intel i9-10850K~CPU @~3.60GHz.
Also,
some groups of orders 1728 and 1920 require quite a few CPUH
to find the irreps.
Orders 768,
1280,
and 1792 have more than one million non-isomorphic groups of each order
and therefore require many CPUH to scan over all of them.

\paragraph{Example:} The discrete group $\mathrm{GL} \left( 2,\, 3 \right)$
has order 48 and {\tt SmallGroups} identifier $\left[ 48,\, 29 \right]$.
By definition,
it is the group generated by three transformations $a$,
$c$,
and $d$ that satisfy~\cite{internet}
\bs
\ba
& & a^4 = c^3 = d^2 = \left( c d \right)^2 = e, \label{1} \\
& & b^2 = a^2, \\
& & b^3 = d a d, \\
& & b a b^{-1} = d b d = a^{-1}, \\
& & b = a^{-1} c a c^{-1}, \label{5}
\ea
\es
where $e$ is the identity transformation and $b \equiv c^{-1} a c$.
There is a faithful representation of these transformations
through $2 \times 2$ unitary matrices:
\be
\label{jgf999}
a \mapsto \frac{1}{3} \left( \begin{array}{cc} i \sqrt{3} & \sqrt{6} \omega \\
  - \sqrt{6} \omega^2 & - i \sqrt{3}
\end{array} \right),
\quad
c \mapsto \left( \begin{array}{cc} \omega & 0 \\ 0 & \omega^2
\end{array} \right),
\quad
d \mapsto \left( \begin{array}{cc} 0 & 1 \\ 1 & 0
\end{array} \right).
\ee
The first two matrices~\eqref{jgf999} have determinant $1$
while the third one has determinant $-1$;
hence,
we classify $\mathrm{GL} \left( 2,\, 3 \right)$ as a subgroup of $U(2)$,
but it is not a subgroup of $SU(2)$.
On the other hand,
there is another faithful irrep of $\mathrm{GL} \left( 2,\, 3 \right)$,
through $4 \times 4$ unitary matrices,
all of them with determinant $1$:
\be
a \mapsto \frac{1}{9} \left( \begin{array}{cccc}
  - 3 \sqrt{3} i & 0 & 6 i & - 3 \sqrt{2} \\
  0 & 3 \sqrt{3} i & 3 \sqrt{2} & - 6 i \\
  6 i & - 3 \sqrt{2} & i \sqrt{3} & - 2 \sqrt{6} \\
  3 \sqrt{2} & - 6 i & 2 \sqrt{6} & - i \sqrt{3}
\end{array} \right),
\quad
d \mapsto \left( \begin{array}{cccc}
  0 & - \omega^2 & 0 & 0 \\ - \omega & 0 & 0 & 0 \\
  0 & 0 & 0 & - 1 \\ 0 & 0 & - 1 & 0
\end{array} \right),
\ee
and $c \mapsto \mathrm{diag} \left( \omega,\ \omega^2,\ 1,\ 1 \right)$.
Therefore,
we classify $\mathrm{GL} \left( 2,\, 3 \right)$
as a subgroup of both $U(2)$ and $SU(4)$,
but $\mathrm{GL} \left( 2,\, 3 \right)$ earns these two classifications
\emph{through different irreps}.

\subsection{The discrete subgroups of $U(3)$ and $SU(3)$}

In Ref.~\cite{jurciukonis}
a classification of the discrete subgroups of $SU(3)$
and of the discrete subgroups of $U(3)$ that are not subgroups of $SU(3)$
has been provided.
All those subgroups were classified
according to their generators and structures.
In this subsection we give relations between the integer $N$
characterizing the center $\zz_N$ of each group
and the identifiers of the group series defined in Ref.~\cite{jurciukonis}.

As explained in subsections~2.2 and~3.1,
the finite subgroups of $SU(3)$
can only have either trivial center or center $\zz_3$;
thus,
they have either $N = 1$ or $N = 3$.
Explicitly,
they have the following values of $N$:
\begin{itemize}
\item The groups $\Delta \left( 3 n^2 \right)$,
  $\Delta \left( 6 n^2 \right)$,
  and $C_{rn,n}^{(k)}$ have $N = 1$ when $n$ cannot be divided by three,
  and $N = 3$ when $n$ is a multiple of three.
\item The groups $D_{3l,l}^{(1)}$ have $N =3$.
\item The exceptional groups
  $\Sigma \left( 60 \right)$ and $\Sigma \left( 168 \right)$ have $N = 1$.
\item The exceptional groups $\Sigma \left( 36 \times 3 \right)$,
  $\Sigma \left( 72 \times 3 \right)$,
  $\Sigma \left( 216 \times 3 \right)$,
  and $\Sigma \left( 360 \times 3 \right)$ have $N = 3$.
\end{itemize}

The series of finite subgroups of $U(3)$ that are not subgroups of $SU(3)$
constructed in Ref.~\cite{jurciukonis} have the following values of $N$:
\begin{itemize}
\item The groups $Y \left( m, j \right)$,
  $L \left( m \right)$,
  $J \left( m \right)$,
  and $\Theta \left( m \right)$ have $N = 3^m$.
\item The groups $T_r^{(k)} \left( m \right)$,
  $\Delta \left( 3 n^2, m \right)$,
  $L_r^{(k)} \left( n, m \right)$,
  $P^{(k)}_r \left( m \right)$,
  $Q^{(k)}_r \left( m \right)$,
  $Q^{(k)\prime}_r \left( m \right)$,
  $X \left( m \right)$,
  $S_r^{(k)} \left( m \right)$,
  $S_r^{(k)\prime} \left( m \right)$,
  $Y_r^{(k)} \left( m \right)$,
  $V_r^{(k)} \left( m \right)$,
  $W \left( n, m \right)$,
  $Z \left( n, m \right)$,
  $Z^\prime \left( n, m \right)$,
  $Z^{\prime \prime} \left( n, m \right)$,
  $\Upsilon \left( m \right)$,
  $\Upsilon^\prime \left( m \right)$,
  and $\Omega \left( m \right)$ have $N = 3^{m-1}$.
\item The groups $M_r^{(k)}$,
  $M_r^{(k)\prime}$,
  $J_r^{(k)}$,
  $Y \left( j \right)$,
  $\tilde Y \left( j \right)$,
  $V \left( j \right)$,
  and $D \left( j \right)$ have $N = 3$.
\item The groups $U \left( n, m, j \right)$ have $N = 3^{j}$.
\item The groups $S_4 \left( j \right)$ have $N = 2^{j-1}$.
\item The groups $\Delta \left( 6 n^2, j \right)$ have $N = 2^{j-1}$
  when $n$ cannot be divided by three,
  and $N = 3 \times 2^{j-1}$ when $n$ is a multiple of three.
\item The groups $\Delta^\prime \left( 6 n^2, m, j \right)$,
  $H \left( n, m, j \right)$,
  $G \left( m, j \right)$,
  $\hat \Xi \left( m, j \right)$,
  and $\Pi \left( m, j \right)$ have $N = 3^{m} 2^{j-1}$.
\item The groups $Z \left( n, m, j \right)$ and $Z' \left( n, m, j \right)$
  have $N = 3^{m-1} 2^{j-1}$.
\item The groups $\Xi \left( m, j \right)$ have $N = 3^{m} 2^{j-2}$.
\end{itemize}
In Ref.~\cite{jurciukonis} a few more subgroups of $U(3)$
that are not subgroups of $SU(3)$ are mentioned,
which could not be classified into any series.
Their values of $N$ are the following:
\begin{itemize}
\item The groups $\left[ 729,\, 96 \right]$,
  $\left[ 729,\, 97 \right]$,
  $\left[ 729,\, 98 \right]$,
  $\left[ 1458,\, 663 \right]$,
  $\left[ 1458,\, 666 \right]$,
  $\left[ 1701,\, 130 \right]$,
  and $\left[ 1701,\, 131 \right]$ have $N = 3$.
\item The group $\left[ 1296,\, 699 \right]$ has $N = 6$.
\item The groups $\left[ 972,\, 170 \right]$,
  $\left[ 1701,\, 102 \right]$,
  and $\left[ 1701,\, 112 \right]$ have $N = 9$.
\end{itemize}

\section{Conclusions}

In this paper we have pointed out that Dark Matter
may be stabilized by a $\zz_N$ cyclic group
under which it has a non-trivial charge---contrary to standard matter,
which is invariant under that $\zz_N$---and that that $\zz_N$
may be the center of the larger internal symmetry group $G$ of Nature,
while $G$ is not a direct product $\zz_N \times G^\prime$.
Thereafter we have performed an extensive
and computationally very time-consuming search for the centers
of discrete groups that cannot be written in the form $\zz_N \times G^\prime$
and that have faithful irreducible representations.
The following are our conclusions:
\begin{itemize}
\item We have found groups with centers $\zz_N$ for $N \le 162$.
\item We have found groups with $N = 2^p \times 3^q$
  for \emph{all} the integers $p$ and $q$ such that $N \le 162$.
\item We have found groups with $N = 2^p \times 5$
  for $0 \le p \le 3$.
\item We have also found groups with $N = 7$,
  $N = 11$,
  $N = 14$,
  $N = 15$,
  and $N = 25$.
\item The number $N$ always divides the order $O$ of the group.
  The integer $O/N$ always has at least two prime factors;
  we have found groups with $O / N = 4$,
  $6$,
  $8$,
  $9$,
  $10$,
  $12$,
  $14$, and so on.
\end{itemize}

In the cases of some smallish groups,
we have explicitly computed the way an element $g$
of order\footnote{The order of an element $g$
is the smallest integer $o$ such that $g^o$ is the unit element of the group.}
$N$ belonging to the $\zz_N$ center of the group
is represented in the various irreps.
We have found that
the sum of the squares of the dimensions of the irreps
where $g$ is represented by any $N^\mathrm{th}$ root of unity
times the unit matrix is always equal to $O/N$.

\paragraph{Acknowledgements:} L.L.\ thanks Salvador Centelles Chuli\'a
for a discussion on his paper~\cite{chulia}.
We thank the {\tt GAP} support team,
specifically Bill Alombert,
Stefan Kohl,
and G\'abor Horv\'ath for helping us solve a problem with that software.
D.J.\ thanks the Lithuanian Academy of Sciences
for financial support through project DaFi2021.
L.L.\ thanks the Portuguese Foundation for Science and Technology
for support through projects UIDB/00777/2020 and UIDP/00777/2020,
and also CERN/FIS-PAR/0004/2019,
CERN/FIS-PAR/0008/2019,
and CERN/FIS-PAR/0002/2021.


\end{document}